\documentclass[12pt]{article} 
\newcommand{\qed}{\hfill \ensuremath{\Box}}
\newtheorem{theorem}{Theorem}
\newtheorem{proposition}{Proposition}
\newtheorem{corollary}{Corollary}
\usepackage{mathtools}
\usepackage{url}
\usepackage{amssymb,amsmath}
\usepackage{graphicx}
\usepackage{subfigure,epsfig}
\title{Coalitional Beliefs of Cournot Network Agents}

\author{Paraskevas V. Lekeas\\
\small Department of Applied Mathematics\\
\small University of Crete, Heraklio, Crete, Greece\\
\small plekeas@gmail.com}

\date{}

\begin{document}

\maketitle

\noindent \textbf{Abstract} In Network cooperative games, due to computational complexity issues, agents are not able to base their behavior on the "whole network status" but have to follow certain "beliefs" as to how it is in their strategic interest to act. This behavior constitutes the main interest of this paper. To this end, we quantify and characterize the set of beliefs that support cooperation of such agents. Assuming that they are engaged in a differentiated Cournot competition and that they equally split the worth produced, we characterize the set of coalitional beliefs that support core non-emptiness and thus guarantee stability of the Network.\\

\noindent \textbf{Keywords} Cooperative game, Cournot competition, Core, Network stability

\section{Introduction}

Game Theory has nowadays become a frequent research tool to approach Network Science. Ideas like Nash equilibrium, price of anarchy and stability, cost allocation,  transferable utilities, coalition formation, core, mechanism design, to name a few, are commonly used not only to understand the behavior of complex network phenomena but also to control and efficiently design such networks. See, for example, Chapter 10 in \cite{MavronicolasSurvay} for a recent survey. Under the above setting, usually there exists an underlying \emph{network} in which some \emph{agents} (or otherwise stated, players, nodes, firms, etc.) are \emph{linked} according to certain relationships and \emph{compete} following specific rules in order to \emph{optimize} a self-interested utility (price, bandwidth, network flow, etc.). However, as many of the real life problems map to large and complex networks \cite{DorogovtsevMendes}, our bounded computational ability and known intractability results \cite{SandhlomLesser,Sandhlometal} restrict our potential to fully unfold and understand coalition formation in networks. 

Due to this complex environment network agents are not able to base their behavior on the "whole network status" but have to follow certain \emph{beliefs} as to how it is in their strategic interest to act. This behavior has gained a lot of attention and such agents are extensively used in Network science \cite{SandhlomLesser,Sandhlometal,Sundararajan,MarginalContribution,Rubinstein,Survey_Jackson,AGT}. These agents constitute the main interest of this paper. To this end, we quantify and characterize the set of beliefs that support cooperation of such agents in a network. In order to illustrate our ideas, we force the agents to compete under a Cournot competition. We also agree that in order to conceive of a group of fully cooperating agents we will use a complete graph (clique). We suppose that at the begining of the competition all the agents, say $N$, are fully cooperating and form a large clique or in game theoretic terms, the \emph{grand coalition}. We then try to answer the following\\

\begin{centering}\fbox {
\parbox{\linewidth}{
\textbf{Basic Question}: \textit{Which are the coalitional beliefs of the agents that support network stability?} 
}}
\end{centering}\\

By network stability we mean the situation in which every agent cooperates with every other agent and no relationship between any two is broken (thus no edge is missing from the graph - network).

In the rest of the paper we proceed as follows: In section 2 we present our model, in sections 3, 4 we present our results and in section 5 we conclude.


\section{The model}

Consider a set of agents $N=\{1,2,\cdots,n\}$ which are the nodes of a graph $G$. Suppose that these agents are engaged in a cooperative game which we will soon define. Whenever two agents $l,m$, $1 \leq l,m \leq n$, cooperate, we draw an edge between
the corresponding nodes in $G$. The graph describing all the cooperations between all the agents in $N$ forms the network. It is clear that when every agent cooperates with every other, then the network is modeled by the complete graph (clique) of $n$ nodes. In this case we say that the
agents form the grand coalition. It is also clear that, unless otherwise stated, every non empty subset of agents $S \subset N$ that fully cooperates also forms a complete subgraph, of $G$, with $s=|S|$ nodes. Let ${\cal{P}}_j$ denote a partition of the \emph{outsiders}, $N \setminus S$, such that $j \in \{1,2,\cdots,n-s\}$ and ${\cal{P}}_j=\{S_{1},S_{2},\cdots,S_j\}$, where $\bigcup\limits_{1 \leq i\leq j} S_i={\cal{P}}_j$ and $S_l \cap S_m = \emptyset$, $\forall \, l \neq m$, $1 \leq l,m \leq j$. We call ${\cal{P}}_j$, a \emph{coalition structure} with $j$ coalitions. Let $s_r=|S_r|$, $r=1,2,\cdots,j$, denote the number of agents in $S_r$. 

The $N$ agents are put together in order to cooperate under a Cournot competition, in which everyone must decide how much "product" to produce, all decisions have to be made at the same time, and each one must take its competitors into account. Each agent $i \in N$, produces with the cost function $C(q_i)=c \cdot q_i$ and with price function $P_i=a-q_i-\gamma\sum\limits_{j=1,j \neq i}^n q_j$, where $q_i$ is the quantity produced. The form of $P_i$ implies that we use product differentiation in the Cournot competition where $a > 0$ is a constant, $c$, $0<c<a$, the per unit cost and $\gamma \in (-1,1]$ a non zero constant. Let $v^{{\cal{P}}_j}_j(S)$ denote the total worth (or value or profit or payoff) of coalition $S$ when $S$ faces coalition structure ${\cal{P}}_j$. Finally, in order to simplify the exposition assume, without loss of generality, that $S=\{1,2,\cdots,s\}$, $S_1=\{s+1,s+2,\cdots,s+s_1\}$, $\cdots$ , $S_j=\{s+\sum\limits_{w=1}^{j-1} s_w+1,\cdots,n\}$.

Consider now the following scenario: a non empty set $S \subset N$ of agents considers not cooperating with the rest and wishes to deviate or in other words disconnect from the initial network (grand coalition). This would result in a split of the network into (at least) two disconnected components, the complete subgraph induced by $S=\{1,2,\cdots,s\}$ and the rest of the network denoted by $N \setminus S$ (the outsiders). Since the consideration of $S$ not to cooperate constitutes a threat for network stability (the grand coalition will break if $S$ manages to deviate), decisions have to be made as to whether specific actions should take place. What would $S$ finally do? The reasonable thing to do is for the agents of $S$ to calculate their worth, a task not trivial at all. This is because their worth depends on how the rest of the network would act. Namely, it depends on the coalition structure that the outsiders $N \setminus S$ will form. 

\section{The worth function}

As stated above, we now consider the scenario where a coalition $S$ contemplates breaking off from the grand coalition. What are the coalitional beliefs of the agents in $S$? In order to answer this we have to reason on what the outsiders, $n-s=|N \setminus S|$, will do. Assume that agents in $S$ believe that the rest will form $j \in \{1,2,\cdots,n-s\}$ coalitions. Since agents compete with each other, everyone faces an optimization problem, the solution to which gives the Nash equilibria describing the worth of each agent in the network when the grand coalition forms. Every agent needs to maximize his worth which is given by the formula: \emph{worth}=(\emph{price-cost})$\times$\emph{quantity produced}. Since agents in every coalition are symmetric, in order to select how much quantity they should produce, they face respectively the following optimization problems:

\[\noindent \overbrace{\max \limits_{(q_i)_{i \in S}} \sum\limits_{i \in S} (a-q_i-\gamma \sum\limits_{l \in N, l \neq i} q_l-c)q_i}^{\text{agents in $S$}}\]

and

\[\overbrace{\max \limits_{(q_h)_{h \in S_k}} \sum\limits_{h \in S_k} (a-q_h-\gamma \sum\limits_{l \in N, l \neq h} q_l-c)q_h}^{\text{outsiders in $S_k$, $k=1,2,\cdots,j$}}\]

One of the ideas that underlies the above problems is that due to symmetry we do not have to solve the same problem for every agent in the same coalition. Instead we can treat every different coalition as a "super agent" and solve the problem concerning him. This approach is common in such problems (see for example \cite{Immorlicaetal}). We have the following\\

\begin{theorem} If coalition $S$ faces structure ${\cal{P}}_j$ and $\gamma > \frac{-1}{n-1}$, then there exist real numbers $\lambda_0,\cdots,\lambda_j$ such that the worth of $S$ is 

\begin{equation}
\label{value final} v^{{\cal{P}}_j}_j(S)=s(1+\gamma s -\gamma)\left(\frac{a-c}{C_0}\right) ^2
\end{equation}

where ${C_0=2[1+\gamma(s-1)]+\gamma
\sum\limits_{k=1}^j\frac{s_k}{\lambda_k}\lambda_0}$.

\end{theorem}

\noindent \textbf{Proof:} For the proof we will need the condition $K : \gamma > \frac{-1}{n-1}$ which guarantees the existence of Cournot equilibrium when $\gamma \in (-1,0)$. The first-order conditions for the optimal quantities of the members of $S$ read

\begin{equation}
2q_i=a-c-2\gamma\sum\limits_{l\in S,l\neq i} q_l-\gamma\sum\limits_{k=s+1}^nq_k, \hspace{0.2cm} i\in S
\nonumber\end{equation}

and for each $S_k$, $k=1,2,...,j$, 

\begin{equation}
2q_h=a-c-2\gamma\sum\limits_{t\in S_k,t\neq h} q_t-\gamma\sum\limits_{r\notin S_k}q_r, \hspace{0.2cm} h\in S_k
\nonumber\end{equation}

Notice that due to (intra-coalitional) symmetry, the following must hold

\begin{equation}
\label{multiplicities}
 \begin{pmatrix}
  q_1=q_2=\cdots=q_s\equiv y_0 \\
  q_{s+1}=q_{s+2}=\cdots=q_{s+s_1}\equiv y_1 \\
  \vdots  \\
   q_{i^*+1}=\cdots=q_n\equiv y_j
 \end{pmatrix}
\end{equation}

where $i^*=s+\sum\limits_{w=1}^{j-1}s_w$. Letting{\footnote{From now on we will use both $s$ and $s_0$ to
denote the number of agents in $S$.}}  $s=s_0,$ the system to solve for is

\begin{equation}
\label{system}
 \begin{pmatrix}
  2y_0=a-c-2\gamma(s_0-1)y_0-\gamma(s_1y_1+s_2y_2+\cdots+s_jy_j) \\
  2y_1=a-c-2\gamma(s_1-1)y_1-\gamma(s_0y_0+s_2y_2+\cdots+s_jy_j) \\
  \vdots  \\
   2y_j=a-c-2\gamma(s_j-1)y_j-\gamma(s_0y_0+s_1y_1+\cdots+s_{j-1}y_{j-1})
 \end{pmatrix}
\end{equation}

Considering $i$th and $k$th equations of (\ref{system}) we have

\begin{align}
2(y_i-y_k)=\gamma (s_k-2)y_k-\gamma(s_i-2)y_i \nonumber \Rightarrow y_k=\frac{y_i(2+\gamma s_i-2\gamma)}{\gamma s_k-2\gamma+2} \\
\label{sumjskyk}\Rightarrow \sum\limits_{k=0,k\neq i}^js_ky_k=\sum\limits_{k=0,k\neq i}^j\frac{s_ky_i(2+\gamma s_i-2\gamma)}{\gamma s_k-2\gamma+2}
\end{align}

Multiplying by $\gamma$, using $k$th equation of (\ref{system}) and the first equation of (\ref{sumjskyk}) we get, 

\begin{equation}
\label{yyii}  y_i=\frac{a-c}{2[1+\gamma(s_i-1)]+\gamma A_i}, \hspace{0.2cm} i=0,1,...,j
\end{equation}

where ${A_i=\sum\limits_{k=0,k\neq i}^j\frac{s_k(2+\gamma s_i-2\gamma)}{\gamma s_k-2\gamma+2}}$. Observe
that $A_i>0$ (due to condition $K$).

Let us now compute the worth of $S$. Recall that

\begin{align}
v_j^{{\cal{P}}_j}(S) &=\sum_{i\in S}(a-q_i-\gamma(\sum\limits_{l\in S,l\neq i}q_l+\sum\limits_{r=s+1}^nq_r)-c)q_i \nonumber \\
\label{compu0} &=s_0(a-y_0-\gamma(s_0-1)y_0-\gamma\sum\limits_{k=1}^js_ky_k-c)y_0
\end{align}

Define $\lambda_k=\gamma s_k-2\gamma+2,$ $k=0,1,...,j$. By using (\ref{sumjskyk}) and (\ref{yyii}) for $i=0$, and the multiplicities of $q$'s for each $y_i$ in (\ref{multiplicities}) we get the following two equations

\begin{align}
&\label{sumjskykallo} \sum\limits_{k=1}^j s_k y_k = \sum\limits_{r=s+1}^n q_r = \sum\limits_{k=1}^j \frac{s_k}{\lambda_k}y_0\lambda_0=\sum\limits_{k=1}^j \frac{s_k}{\lambda_k}\frac{a-c}{C_0}\lambda_0\\
&\label{sumy0} \sum\limits_{l \in S \setminus \{0\}} y_0=\sum \limits_{l \in S \setminus \{0\}}q_l=(s_0-1)\frac{a-c}{C_0}
\end{align}

where ${C_0=2[1+\gamma(s_0-1)]+\gamma \sum\limits_{k=1}^j\frac{s_k}{\lambda_k}\lambda_0}$. Using (\ref{compu0},\ref{sumjskykallo},\ref{sumy0}) we get the theorem.  \qed

~\

Using Theorem 1 we have the following\\

\begin{corollary} When the grand coalition forms, the total worth of the network is
\begin{equation}
v(N)=\frac{n(a-c)^2}{4(1+\gamma(n-1))} \; \text{.}
\nonumber\end{equation}
\end{corollary}

Corollary 1 can be very useful in the following sense. Since, upon cooperation, the agents split the profit equally, when the grand coalition forms each of them will receive $\frac{v(N)}{n}$. This means that no agent would accept deviating from the grand coalition and join a smaller one if he did not first ensure that he would receive at least more than $\frac{v(N)}{n}$ with this move. So a first approach to the Basic Question is the following\\

\begin{centering}\fbox {
\parbox{\linewidth}{
\textbf{Answer 1:} \textit{If no agent assures a profit of at least more than $\frac{(a-c)^2}{4(1+\gamma(n-1))}$, then no agent likes to deviate and thus the grand coalition forms and the network does not break.}}}
\end{centering}\\

When the grand coalition forms it is said that the core of the game is non-empty and since we would like the whole network to be stable and not to break into disconnected components, we have to characterize the set of coalitional beliefs that support core non-emptiness. So it is of special interest to study the core of the above game. This is the subject of the following section.

\section{The core}

Before saying anything about the core of the game, let us examine the impact of the distribution of $(s_1,\cdots,s_j)$ on $v^{{\cal{P}}_j}_j(S)$. This is useful in case agents in $S$ need to follow a trend in their reasoning. When $\gamma \in (0,1)$ we have the following\\

\begin{proposition} For a fixed $j$, the optima of $v^{{\cal{P}}_j}_j(S)$, w.r.t. ($s_1,\cdots,s_j$), are

\noindent (i) min when $s_i=\frac{n-s}{j},1\leq i \leq j$

\noindent (ii) max when $s_i=1$, in all but one value $h$, and $s_h=n-s-(j-1)$, $h \in \{1,\cdots,j\}$ .

\end{proposition}

\noindent \textbf{Proof:} From (\ref{value final}), $v^{{\cal{P}}_j}_j(S)$ is min when $C_0$ is max which happens when $\sum\limits_{k=1}^j \frac{s_k}{\lambda_k}$ is max which happens when $s_i=\frac{n-s}{j},1\leq i \leq j$, because

\begin{align}
\sum\limits_{k=1}^j\frac{1}{\gamma s_k-2\gamma+2}=
&\sum\limits_{k=1}^{j-1}\frac{1}{\gamma s_k-2\gamma+2}+\frac{1}{\gamma s_j-2\gamma+2} \nonumber \\
=\label{max1} &\sum\limits_{k=1}^{j-1}\frac{1}{\gamma s_k-2\gamma+2}+\frac{1}{\gamma (n-s-\sum\limits_{k=1}^{j-1}s_k)-2\gamma+2}
\end{align}

Let $h(s_1,s_2,\cdots,s_{j-1})$ denote the right hand side of (\ref{max1}). To find the vector of $(s_1,...,s_{j-1})$ that minimizes (\ref{max1}) we need to solve the system

\begin{equation}
\frac{\partial{h(s_1,s_2,...,s_{j-1})}}{\partial{s_k}}=0, \hspace{0.2cm}k=1,2,\cdots,j-1
\end{equation}

Straightforward calculations lead to the solution $s_1=\cdots=s_{j-1}=\frac{n-s}{j}$ (and hence $s_j=\frac{n-s}{j}$), so we have (i). In a similar way we can also prove (ii). \qed

~\

Proposition 1 gives the deviating agents a first belief about their potential worth\\

\begin{centering}\fbox {
\parbox{\linewidth}{
\textbf{Answer 2:} If the number of coalitions that outsiders form is fixed, say $j$, then the worth of the deviating agents, $s$, is minimized (hence the network is less likely to break) when the $n-s$ agents split equally among the $j$ coalitions; and the worth of $s$ is maximized when $j-1$ coalitions have one member and one coalition has $n-s-(j-1)$ members.}}
\end{centering}\\

The above answer gives in a way an incentive to the deviating agents, meaning that a coalition of agents has a lower incentive to deviate when its opponents are split equally among $j$ (see Fig. \ref{fig:MaxMin}). 

Another look at how agents in $S$ might reason is given by the following

\begin{proposition} When $\gamma = 1$, if a coalition $S$ believes that the $n-s$ outsiders will form more than $j^*(n,s)=2(\sqrt{\frac{n}{s}}-1)$ coalitions, then $S$ will not deviate from the grand coalition.
\end{proposition}

\noindent \textbf{Proof:} Before giving the proof we make the following observation. As said earlier, since the agents upon cooperation split the profit equally, they know that when the network does not break they are going to receive $\frac{v(N)}{n}$ each. On the other hand, agents belonging to a deviating coalition $S$, after calculating their worth, know that upon deviation they are going to receive $\frac{v(S)}{s}$ each. So if $\frac{v(N)}{n} \geq \frac{v(S)}{s}$, then no agent in $S$ would like to deviate. Now returning to the proof, from (\ref{yyii}) for $\gamma=1$ we get $y_i=\frac{a-c}{s_i(2+j)}$, $0 \leq i \leq j$. Hence the value of $S$ is independent of the size of coalitions of the outsiders but depends only on their number, i.e. $v^{{\cal{P}}_j}_j(S)=\left(\frac{a-c}{j+2}\right)^2$. The core is then non empty if for all $S$, the inequality $\frac{v(N)}{n}\geq \frac{v(S)}{s}$ holds, or if $\frac{1}{4n} \geq \frac{1}{s(j+2)^2}$. The last inequality holds iff $j>j^*(n,s)=2(\sqrt{\frac{n}{s}}-1)$. \qed

~\

Proposition 2 implicitly provides agents with another belief in the special case where $\gamma=1$\\

\begin{centering}\fbox {
\parbox{\linewidth}{
\textbf{Answer 3:} \textit{In the special case where $\gamma=1$, agents do not have to worry about what the rest will do, meaning that they do not have to worry about how the rest are going to split. They only have to worry about how many of them there are, i.e. their total number.}
}
}
\end{centering}\\

A more thorough look at how agents in $S$ might think is given by the following

\begin{theorem} Assume $\gamma \in (0,1]$. There exists a $j^*(n,s,\gamma)$, such that the core is non-empty for all $j>j^*(n,s,\gamma)$ and all corresponding ${\cal{P}}_j$.
\end{theorem}

\noindent \textbf{Proof:}  We seek conditions for the inequality $\frac{v(N)}{n}\geq \frac{v_j^{{\cal{P}}_j}(S)}{s}$ to hold:

\begin{align}
&C_0 \geq 2\sqrt{(1+\gamma n-\gamma)(1+\gamma s-\gamma)} \nonumber \\
\text{or} \; &2(1+\gamma s-\gamma)+\gamma \sum_{k=1}^j \frac{s_k}{\lambda_k}\lambda_0 \geq 2\sqrt{(1+\gamma n-\gamma)(1+\gamma s-\gamma)} \nonumber\\ 
\text{or} \; &\sum_{k=1}^j \frac{\gamma s_k}{\gamma s_k+2(1-\gamma)} \geq 2\left(\sqrt{\frac{1+\gamma n-\gamma}{1+\gamma s-\gamma}}-1\right)/(1+\frac{1-\gamma}{1+\gamma s -\gamma}) \equiv \zeta \nonumber
\end{align} 

Since $0 < \gamma \leq 1$ the sum can not exceed $j$, so for $j^*(n,s,\gamma)=\zeta$ the core is non empty for all $j>j^*(n,s,\gamma)$.

We now have to show that $n-s > \zeta$. But it can be easily proven by induction on $s, 0<s<n$ that for fixed $n, n>1$ and $\gamma \in (0,1]$, $\zeta < n-s$. And since for every $s<n$, $\zeta > 0$,  we have the theorem.  \qed

~\

Theorem 2 gives, in a computational efficient way, a belief as to how agents in $S$ might think when considering deviation from the Network and thus constitutes another answer to our Basic Question (see also Fig.  \ref{fig:MaxMin}):\\

\begin{centering}\fbox {
\parbox{\linewidth}{
\textbf{Answer 4:} \textit{When $\gamma \in (0,1]$, the network does not break provided that agents in $S$ believe the outsiders will form a structure with a sufficiently large number of coalitions ($>j^*$)}.}
}
\end{centering}\\

Using the formula for $j^*(n,s,\gamma)$ we can prove proposition 2 in a different way by plugging $\gamma=1$ and get $j^*(n,s,1)=\zeta|_{\gamma=1}=2(\sqrt{\frac{n}{s}}-1)$ as expected. 

Another perspective of the beliefs of agents in $S$ is given by the following

\begin{theorem}
If $\gamma \in (-1,0)$ and condition $K$ holds, then the core is non-empty for all $j=1,2,\cdots,n-s$ and all ${\cal{P}}_j$.
\end{theorem}

\noindent \textbf{Proof:} The inequality $\frac{v(N)}{n}\geq \frac{v_j^{{\cal{P}}_j}(S)}{s}$ holds iff

\begin{equation}
2+\gamma(1+\frac{1-\gamma}{1+\gamma s -\gamma}) \sum_{k=1}^j \frac{s_k}{\lambda_k} \geq 2\frac{{\sqrt{(1+\gamma n-\gamma)(1+\gamma s-\gamma)}}}{{(1+\gamma s-\gamma)}}
\nonumber\end{equation}

or using the definitions $\sigma \equiv 1+\gamma(s-1)$,  $\nu \equiv 1+\gamma(n-1)$, it holds iff

\begin{align}
2+\frac{\gamma \lambda_0}{\sigma} \sum_{k=1}^j \frac{s_k}{\lambda_k} \geq 2\sqrt{\frac{\nu}{\sigma}}
\; \; \; \text{or} \; \; \;
\gamma \geq \frac{2(\sqrt{\nu}-\sqrt{\sigma})\sqrt{\sigma}}{\lambda_0\sum_{k=1}^j \frac{s_k}{\lambda_k}}
\nonumber\end{align}

Since $\gamma > \frac{-1}{n-1}$, it suffices to show that

\begin{align}
\frac{-1}{n-1}>\frac{2(\sqrt{\nu}-\sqrt{\sigma})\sqrt{\sigma}}{\lambda_0\sum_{k=1}^j \frac{s_k}{\lambda_k}}
\; \; \; \text{or} \; \; \;
\frac{2(n-1)(\sigma-\sqrt{\sigma \nu})}{\lambda_0} > \sum \limits_{k=1}^{j} \frac{s_k}{\lambda_k}
\nonumber\end{align}

The maximum value of the term ${\sum \limits_{k=1}^{j} \frac{s_k}{\lambda_k}}$ appears at
$s_h=1$ for all but one $h$ (remember that now $\gamma<0$). It suffices then to show that the left hand side of the last inequality is greater than the corresponding maximum value of the sum. So we must show

\begin{equation}
\frac{2(n-1)(\sigma-\sqrt{\sigma \nu)}}{\lambda_0} > \frac{j-1}{-\gamma+2}+\frac{n-s-(j-1)}{\gamma(n-s-(j-1))-2\gamma+2}
\nonumber\end{equation}

\begin{align} \text{But} \; \;  \lambda_0=\gamma s -2\gamma + 2 < 2 - 2\gamma \; \; \text{so} \; \; \frac{2(n-1)(\sigma-\sqrt{\sigma \nu})}{\lambda_0}>\frac{2(n-1)(\sigma-\sqrt{\sigma \nu})}{2-2\gamma}
\nonumber\end{align}

\begin{align} \text{or that} \; \; \label{++}\frac{{2}(n-1)(\sigma-\sqrt{\sigma \nu)}}{{2}(1 -\gamma)} > \frac{j-1}{2-\gamma}  + \frac{n-s-(j-1)}{\gamma(n-s-(j-1))-2\gamma+2}
\end{align}

Since $\frac{j-1}{1-\gamma}>\frac{j-1}{2-\gamma}$, instead of inequality (\ref{++}) we can prove that

\begin{equation}
\label{A+} \frac{(n-1)(\sigma-\sqrt{\sigma \nu)}-(j-1)}{1 -\gamma} > \frac{n-s-(j-1)}{\gamma(n-s-(j-1))-2\gamma+2}
\end{equation}

\begin{align}
\text{But} \; \; \gamma(n-s-(j-1))>-1 \; \; \text{so} \; \; \frac{n-s-(j-1)}{-1-2\gamma+2} > \frac{n-s-(j-1)}{\gamma(n-s-(j-1))-2\gamma+2}
\nonumber\end{align}

\begin{align}
\text{Instead of (\ref{A+}) we can prove} \; \; &\frac{(n-1)(\sigma-\sqrt{\sigma \nu)}-(j-1)}{1 -\gamma} > \frac{n-s-(j-1)}{1-2\gamma}\nonumber\\
\text{or} \; \; \; &\frac{(n-1)(\sigma-\sqrt{\sigma \nu)}-{(j-1)}}{{1 -\gamma}} > \frac{n-s-{(j-1)}}{{1-\gamma}}\nonumber\\
\text{because} \; \; \; &\frac{n-s-(j-1)}{1-\gamma} > \frac{n-s-(j-1)}{1-2\gamma}\nonumber
\end{align}

\begin{align}
\text{So must show that} \; \; &\sigma-\sqrt{\sigma \nu} > \frac{n-s}{n-1}=\frac{n-s+1-1}{n-1}=1-\frac{s-1}{n-1}\nonumber\\
\text{or} \; \; &1+\gamma(s-1)-\gamma\sqrt{(\frac{1}{\gamma}+(s-1))(\frac{1}{\gamma}+(n-1))} > 1-\frac{s-1}{n-1}\nonumber\\
\text{or} \; \; &\gamma(s-1)-\gamma\sqrt{(\frac{1}{\gamma}+(s-1))(\frac{1}{\gamma}+(n-1))} > -\frac{s-1}{n-1}\nonumber
\end{align}

Since $\gamma (s-1) > -\frac{s-1}{n-1}$, it suffices to show that

\begin{align}
\gamma(s-1)-\gamma\sqrt{(\frac{1}{\gamma}+(s-1))(\frac{1}{\gamma}+(n-1))} > \gamma(s-1)  \; \; \text{which holds. \qed}
\nonumber\end{align}

Theorem 3 gives us another answer to the Basic Question (see also Fig. \ref{fig:j*}):\\

\begin{centering}\fbox {
\parbox{\linewidth}{
\textbf{Answer 5:} \textit{The network never breaks when $\gamma \in (-1,0)$ and $\gamma > \frac{-1}{n-1}$, so no group of agents will consider deviating from the grand coalition under these conditions.}}
}
\end{centering}

\section{Conclusions}

In this paper we forced a set of $n$ agents to compete under a differentiated Cournot competition and by imagining their full cooperation as a complete graph we studied the various beliefs that support core non-emptiness and thus network stability. Fixing the number of coalitions that outsiders form to say $j$ coalitions, we proved that the worth of the deviating agents, $s$, is minimized (hence the network is less likely to break) when the $n-s$ agents split equally among the $j$ coalitions; and the worth of $s$ is maximized when $j-1$ coalitions have one member and one coalition has $n-s-(j-1)$ members. Given the above, we proved that when $\gamma > 0$ the network does not break, provided the agents of a deviant coalition believe their opponents will form a sufficiently large number of coalitions. On the other hand, if $\gamma < 0$ and $\gamma > \frac{-1}{n-1}$, the network does not break irrespective of the beliefs of the deviant coalition, so no agent will consider deviating. Finally, we proved that when $\gamma=1$, the deviating agents, in order to calculate their worth and thus decide what to do, must only worry about how many coalitions are going to be formed by the outsiders. 

\subsection*{Acknowledgements} I am greatly indebted to Professor Giorgos Stamatopoulos for advising and encouraging me to finish this work.

\newpage 

\section*{Appendix}
\begin{center}
\includegraphics[scale=0.65]{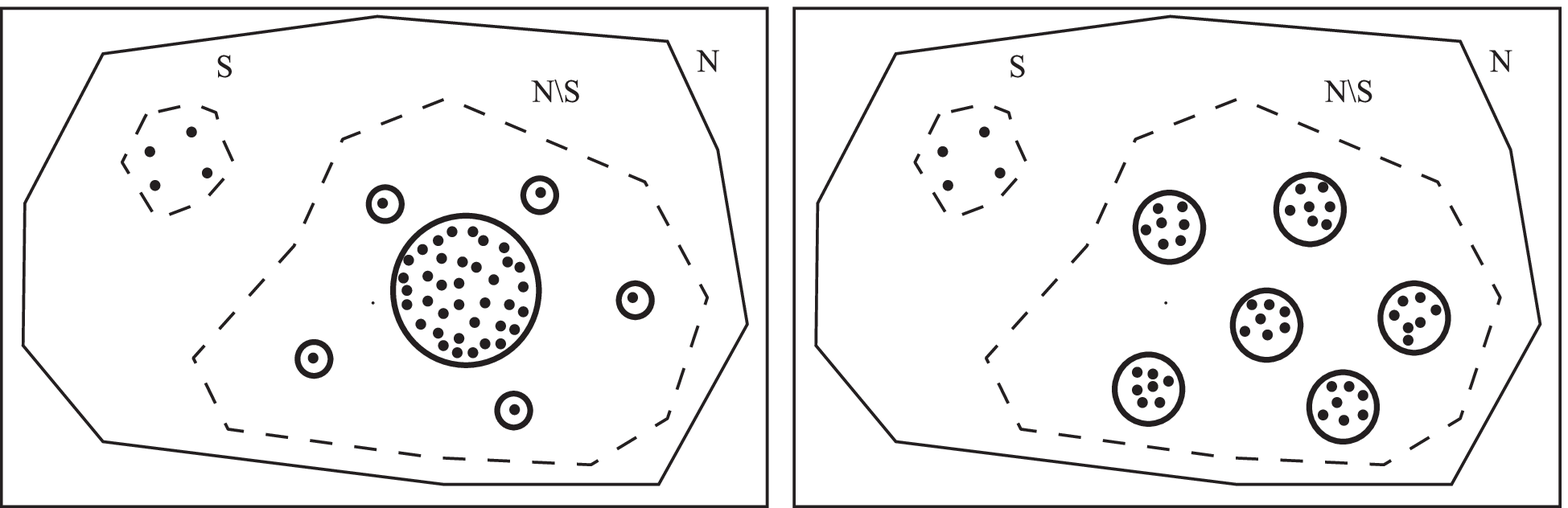}
\end{center}
\begin{figure}[htbp]
\caption{On the left, when $\gamma \in (0,1)$ and for a fixed $j$, the worth of $S$ is maximized when the outsiders, $N \setminus S$, split in all coalitions but one into singletons. In this figure we have the case where $n=46$, $s=4$, and $j=6$. Respectively, on the right the worth is minimized when they split equally among the $j=6$ coalitions, having $\frac{n-s}{j}=7$ agents each.} \label{fig:MaxMin}
\end{figure}
\begin{center}
\includegraphics[scale=0.65]{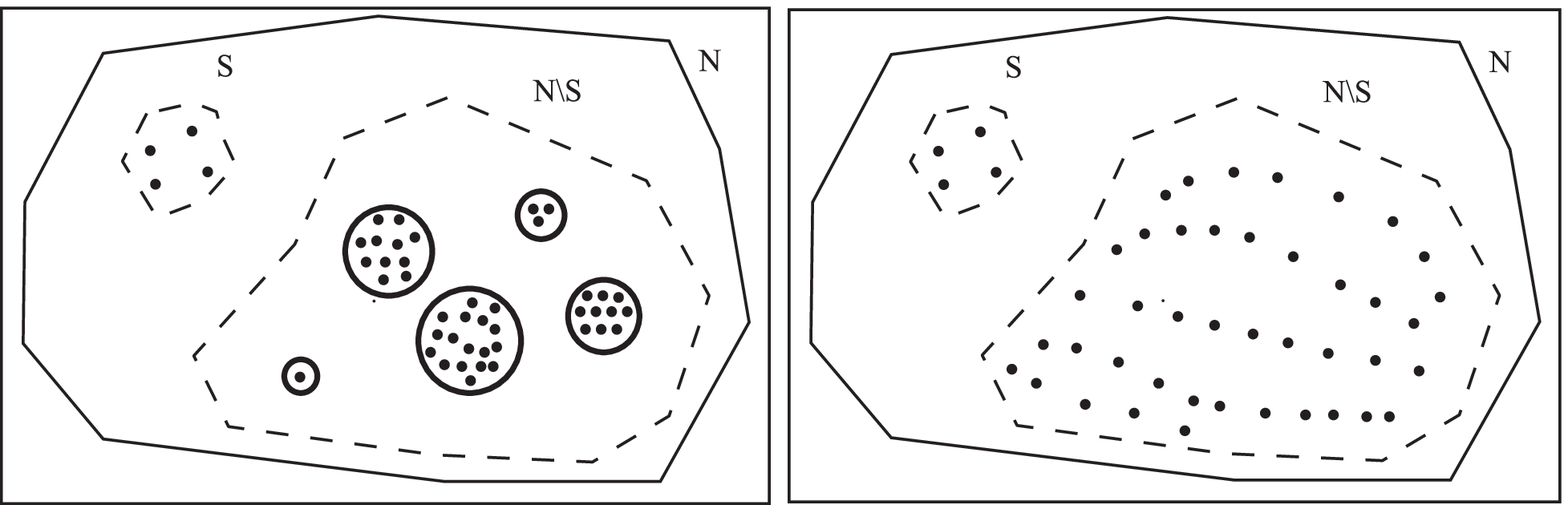}
\end{center}
\begin{figure}[htbp]
\caption{On the left, when $\gamma \in (0,1]$ the network does not break provided that agents in $S$ believe the outsiders will form a large $>j^*$ number of coalitions. In this figure we illustrate the case where $n=46$, $s=4$, and $\gamma=0.9$, for which $j^*(46,4,0.9) \approx 4.57$. So when the outsiders split into 5 coalitions the network does not break. On the right, when $\gamma \in (-1,0)$, the network does not break irrespective of what $S$ might think.} \label{fig:j*}
\end{figure}


\newpage

\begin{thebibliography}{1}

\bibitem{MavronicolasSurvay} 
Mavronicolas, M., Papadopoulou, V. and Spirakis, P. Algorithmic Game Theory and Applications. In {\em Handbook of Applied Algorithms: Solving Scientific, Engineering, and Practical Problems}; N., Nayak, I., Stojmenovic Eds; Wiley-IEEE Press, 2008.

\bibitem{DorogovtsevMendes}
Dorogovtsev, S. N., Mendes, J. F. F. Evolution of networks. {\em Advances in Physics} {\bf 2002}, {\em 51}, 4, Taylor and Francis (2002)

\bibitem{SandhlomLesser} 
Sandhlom T., Lesser V. Coalitions among computationally bounded agents. {\em Artificial Intelligence (Special issue on Economic Principles of Multi-Agent Systems)}, {\bf 1997}, {\em 94}, 1-2, 99-137.

\bibitem{Sandhlometal}
Sandholm, T., Larson, K., Andersson M., Shehory O., and Tohme F. Coalition structure generation with worst case guarantees. {\em Artificial Intelligence} {\bf 1999}, {\em 111}, 1-2, 209-238. 

\bibitem{Sundararajan} 
Sundararajan A. Local Network Effects and Complex Network Structure. {\em The B.E. Journal of Theoretical Economics} {\bf 2008}, {\em 7}

\bibitem{MarginalContribution} 
Kirman, A., Markose, S., Giansante, S., and Pin, P. Marginal contribution, reciprocity and equity in segregated groups: Bounded rationality and self-organization in social networks. {\em Journal of Economic Dynamics and Control} {\bf 2007}, {\em 31}.

\bibitem{Rubinstein}
Rubinstein, A. Modeling Bounded Rationality, MIT Press, 1997

\bibitem{Survey_Jackson} 
Jackson, K A Survey of Models of Network Formation: Stability and Efficiency. Group Formation in Economics: Networks, Clubs, and Coalitions, Cambridge University Press, 2005

\bibitem{AGT} 
Nisan, N., Roughgarden, T. , Tardos, E. and Vazirani, V. Algorithmic Game Theory, Cambridge University Press 2007  

\bibitem{Immorlicaetal} 
Immorlica N., Markakis E. and Piliouras G. oalition Formation and Price of Anarchy in Cournot Oligopolies. In {\em 6th Workshop on Internet and Network Economics}, LNCS, 2010, Volume 6484/2010, 270-281

\end{thebibliography}
\end{document}